\documentclass[aps,prd,twocolumn,showpacs,groupedaddress,floatfix,nofootinbib,letterpaper]{revtex4}

\usepackage{amsmath,amsthm,amssymb,enumerate}
\usepackage{graphicx}
\usepackage{hyperref}
\usepackage{subfig}

\newcommand{\abs}[1]{\left\vert#1\right\vert}
\newcommand{\avg}[1]{\left\langle#1\right\rangle}

\def\beq{\begin{equation}}
\def\eeq{\end{equation}}

\def\tvarphi{\tilde{\varphi}}

\begin{document}

\title{Coupling the inflaton to an expanding aether}
\author{William Donnelly}
\email{wdonnell@umd.edu}
\author{Ted Jacobson}
\email{jacobson@umd.edu}
\affiliation{
Center for Fundamental Physics, Department of Physics \\
University of Maryland, College Park, MD 20742-4111
}

\begin{abstract}
We consider a Lorentz-violating theory of inflation consisting of Einstein-aether theory with a scalar inflaton coupled bilinearly to the expansion of the aether.
We determine the conditions for linearized stability, positive energy and vanishing of preferred-frame post-Newtonian parameters, and
find that all these conditions can be met.
In homogeneous and isotropic cosmology, the inflaton-aether 
expansion coupling leads to a  
driving force on the inflaton that 
is proportional to the Hubble parameter.
This force
affects the slow-roll dynamics, but still allows
for a natural end to inflation.
\end{abstract}


\maketitle

\section{Introduction} \label{section:introduction}

The space-time continuum may be only a low-resolution approximation to a more structured, discrete plenum, with a UV cutoff on degrees of freedom. This structure might be revealed in high energy collisions or subtle effects of long distance propagation, or discrete symmetry violations. But it might also manifest itself at a macroscopic level as a result of the expansion of the Universe.  The expansion would presumably require that new ``atoms of space-time" together with new degrees of freedom are created as time goes on. The aim of this paper is to explore possible consequences of this scenario using an effective field theory description. 

A UV cutoff is inconsistent with locality and Lorentz symmetry, so at least one of these properties would have to fail if such a cutoff exists. In this paper we study a model theory in which locality is preserved, but Lorentz symmetry fails. To accommodate this scenario in an effective field theory that preserves general covariance and the successes of general relativity, one needs to incorporate a Lorentz-violating dynamical field into the theory. Einstein-aether theory provides a straightforward approach to doing this. In that theory, the local structure of space-time is described by a metric tensor, as in general relativity, but also by a dynamical unit timelike vector field $u^a$, called the aether. The aether defines a preferred rest frame at each point of space-time, but preserves rotational symmetry in that frame. This structure would suffice to accommodate a local UV cutoff.
For a Lorentz violating theory like this to be viable, there
must ultimately be a reason why conspicuous
Lorentz violation does not infect the low-energy matter action
in flat space-time.
For the purposes of this paper, 
we will simply assume that such a reason exists.

Just as the Einstein-Hilbert action is (besides a cosmological constant term) the unique lowest order covariant term in a derivative expansion of the action, the action for ``pure" Einstein-aether theory consists, 
in addition,
of the four independent two-derivative terms involving the metric and the aether. Direct coupling of the aether to matter would entail local Lorentz violation and is severely constrained by observational bounds \cite{Mattingly2005}.

Our primary goal is to use Einstein-aether theory to model possible consequences 
associated with the growth of the vacuum in cosmology. 
In particular, we study the leading order coupling of a scalar inflaton field $\varphi$ 
to the local expansion rate. 
Such a term cannot appear in a purely metric theory because the local expansion rate, 
i.e. the Hubble parameter $H$, cannot be constructed from the metric in a covariant way. 
However, given the aether field $u^a$, the local expansion $\theta$ relative to the 
preferred frame is a space-time scalar,
\begin{equation}
\theta \equiv \nabla_a u^a.
\end{equation}
(In a homogeneous and isotropic cosmology, $\theta$ is related to the Hubble parameter by $\theta = 3 H$.) 
Any scalar operator might couple to $\theta$, but here we shall consider just 
a (still mysterious) scalar inflaton, since that would presumably dominate during inflation.
This of course leaves open the possibility that scalar operators composed of 
matter fields could couple to $\theta$. 
At late times $\theta$ is everywhere small relative to particle physics scales, so $\theta$ couplings would not lead to any observable Lorentz violation in particle physics.

We shall consider a generic coupling of the 
inflaton to the expansion, but the lowest dimension 
(and therefore presumably dominant) coupling takes the form 
\beq
\theta\varphi=-u^a\nabla_a\varphi + \mbox{total derivative}. 
\eeq
This peculiar leading order term has not, as far as we know, 
been considered previously. It apparently entails a violation of 
time-reversal symmetry, although if the rest of the inflaton action is 
even in $\varphi$ then time-reversal symmetry is preserved 
when accompanied with 
$\varphi \to -\varphi$.  We first examine the impact of this
new coupling on all of the noncosmological theoretical and 
observational constraints on the theory, and then we study 
the modified inflationary dynamics. A key consequence of this 
term is that the expansion $\theta$ acts as an external force, 
which can either slow down or accelerate the evolution of the inflaton. 
Its effect on the primordial fluctuations has not yet been fully investigated. 

Although the particular coupling we examine may be new, 
much work has already been done on cosmology in Einstein-aether theory. 
Inflation with a preferred timelike vector was 
first considered by Gasperini \cite{Gasperini1985}. 
More recently, authors have computed the effect of 
the aether on the spectrum of primordial perturbations 
\cite{Lim2004, Li2007, Zuntz2008, ArmendarizPicon2010}, 
with no direct coupling between the aether and the inflaton. 
A certain kind of aether-inflaton coupling has 
been considered in which the kinetic terms in the 
aether action are multiplied by functions of the 
scalar field \cite{Kanno2006, Avelino2009}. 
In this model the effective gravitational constant 
depends on the inflaton, leading to the possibility 
of a repulsive phase that significantly changes the 
dynamics of inflation.
A similar model with a nonnormalized vector 
field playing the role of the aether and scalar 
fields has been proposed as a solution of the 
cosmological constant problem by a similar mechanism
\cite{Klinkhamer2009}. 
Cosmology with a Lagrangian that is an arbitrary 
function of the usual Einstein-aether Lagrangian 
has also been considered as a possible source of dark energy 
\cite{Zlosnik2006, Zuntz2010}.

This paper is organized as follows. In Sec.~\ref{section:action} the theory is 
defined and the equations of motion are given. In Secs.~\ref{section:modes}, \ref{section:stability}, and \ref{section:energy} we work out the properties of linearized spin-0 perturbations
and the conditions for their stability, absence of 
vacuum Cherenkov radiation, and energy positivity.
This is necessary in order to establish the conditions for viability of the 
model with aether-scalar coupling. 
In Sec.~\ref{section:constraints} these conditions are combined with those for 
the spin-1 and spin-2 perturbations of pure Einstein-aether theory,
the effect of the new coupling on the 
parametrized post-Newtonian (PPN) expansion is deduced, and some additional observational constraints are discussed.
In Sec.~\ref{section:cosmology} the homogeneous, isotropic cosmological dynamics
is investigated, and we conclude with a summary and some further 
remarks in Sec.~\ref{section:conclusion}. 

\section{Action and equations of motion} \label{section:action}

Einstein-aether theory is a theory of a metric $g_{ab}$ of signature $(+,-,-,-)$ and a unit timelike vector field $u^a$ \cite{Jacobson2008}.
This theory is defined by the Lagrangian
\begin{equation} \label{Lae}
L_{\text \ae} = \frac{-M^2}{2} \left( R + K^{ab}_{\phantom{ab}cd} \nabla_a u^c \nabla_b u^d + \lambda(u^a u_a - 1) \right),
\end{equation}
where $M \equiv (8 \pi G)^{-\frac{1}{2}}$ is the reduced Planck mass. 
In general relativity, the sign of $G$ can be fixed by requiring that 
gravitational waves carry positive energy density. The same 
turns out to be true in Einstein-aether theory, hence we 
will assume from the outset that $G>0$, or equivalently
$M^2>0$.  
The three terms in the action are the Ricci scalar 
for $g_{ab}$, the Einstein-aether kinetic term, 
and a Lagrange multiplier term that forces $u^a$ to be a timelike unit vector. 
The tensor $K^{ab}_{\phantom{ab}cd}$ is given by
\begin{equation}
K^{ab}_{\phantom{ab}cd} = c_1 g^{ab} g_{cd} + c_2 \delta^a_c \delta^b_d + c_3 \delta^a_d \delta^b_c + c_4 u^a u^b g_{cd},
\end{equation}
where $c_{1,2,3,4}$ are the dimensionless free parameters of Einstein-aether theory. 

The theory defined by the Lagrangian (\ref{Lae}) has a Newtonian limit with a modified Newton's constant
\begin{equation} \label{G_n}
G_N = \kappa_N G, \qquad \kappa_N = \frac{2}{2 - c_{14}}.
\end{equation}
The post-Newtonian expansion will be discussed in Sec.~\ref{section:ppn}.
The Friedmann equation 
for spatially flat homogeneous cosmologies 
is also the same as in general relativity, but with a renormalized Newton's constant $G_C$ in place of $G$ \cite{Mattingly2001,Carroll2004},
\begin{equation} \label{G_c}
G_C = \kappa_C G, \qquad \kappa_C = \frac{2}{2 + c_{13} + 3 c_2}.
\end{equation}
The spatial curvature term in the Friedmann equation is also renormalized.

We introduce to the theory a scalar inflaton $\varphi$ with a potential that depends both on $\varphi$ and the expansion of the aether,
\begin{equation} \label{Lphi}
L_\varphi = \frac{1}{2} \nabla_a \varphi \nabla^a  \varphi - V(\theta, \varphi).
\end{equation}
For special cases of the potential $V(\theta, \varphi)$ the cosmological dynamics of our model overlaps with other models of Lorentz-violating inflation. To see this, we note that in a Friedmann-Lema\^itre-Robertson-Walker metric with the aether aligned to the cosmological rest frame, 
\begin{equation}
\nabla_a u_b = H(g_{ab} - u_a u_b),
\end{equation}
where $H = \dot{a}/a$ is the Hubble parameter. It follows that the aether kinetic term in (\ref{Lae}) is proportional to $\theta^2$. 
In the model of Kanno and Soda \cite{Kanno2006}, 
$c_{1,2,3,4}$ are allowed to depend on a scalar field $\varphi$.
This will have the same cosmological dynamics as our model with a potential of the form $V(\theta, \varphi) = f(\varphi) \theta^2$.
The model of Zlosnik et.\ al.\ \cite{Zlosnik2006}
does not have a scalar field, but 
its Lagrangian contains an arbitrary function of the aether kinetic term.
This cosmology corresponds to a potential 
of the form $V(\theta, \varphi) = f(\theta^2)$.
This equivalence holds only for the background dynamics; the linearized perturbations of these models will generally differ.

The equation of motion for $\varphi$ is
\begin{equation} \label{klein-gordon}
\square \varphi + V_\varphi = 0,
\end{equation}
where $V_\varphi$ denotes the partial derivative $\partial V/\partial\varphi$.
The equation of motion for $u$ is
\begin{equation} \label{u}
\nabla_b K^b_{\phantom{b}a} = \lambda u_a + c_4(u^c \nabla_c u_b) \nabla_a u^b - M^{-2} \nabla_a V_\theta ,
\end{equation}
where $K^b_{\phantom{b}d} \equiv K^{ab}_{\phantom{ab}cd} \nabla_a u^c$.
The component of  Eq.  (\ref{u}) parallel to $u^a$ determines $\lambda$. 
Finally we have the Einstein equation:
\begin{equation}
G_{ab} = S_{ab} + M^{-2} T_{ab},
\end{equation}
where $S_{ab}$ is the energy-momentum tensor for the aether in the absence of the scalar field \cite{Eling2006}
\begin{align} \label{S}
S_{ab} &= \nabla_c \left( K_{(a}^{\phantom{(a}c} u^{\phantom{(ac}}_{b)} - K^c_{\phantom{c}(a} u^{\phantom{c}}_{b)} - K_{(ab)}u^c \right) \nonumber \\
&\quad + c_1 (\nabla_c u_a \nabla^c u_b - \nabla_a u_c \nabla_b u^c) \nonumber \\
&\quad+ c_4 (u^c \nabla_c u_a) (u^d \nabla_d u_b) \nonumber \\
&\quad+ \left(u^c \nabla_d K^d_{\hphantom{d}c} - c_4 (u^c \nabla_c u^e)(u^d \nabla_d u_e) \right) u_a u_b \nonumber \\
&\quad+ \frac{1}{2} g_{ab} (K^c_{\phantom{c}d}\nabla_c u^d),
\end{align}
round brackets denoting symmetrization.
The matter energy momentum tensor $T_{ab}$
is obtained by varying the matter action with respect to the metric. 
Accounting for the contribution of the Lagrange multiplier term
the energy-momentum tensor can be written in terms of the Lagrangian as
\begin{equation}
T_{ab} = 2 \frac{\delta L}{\delta g^{ab}} + u^c \frac{\delta L}{\delta u^c} u_a u_b - L g_{ab}.
\end{equation}
For the scalar Lagrangian (\ref{Lphi}) this formula gives
\begin{align}\label{scalar-stress}
T_{ab} &= \nabla_a \varphi \nabla_b \varphi - \left( \tfrac{1}{2} \nabla_a \varphi \nabla^a \varphi - V  + \theta V_\theta \right) g_{ab}\nonumber \\
&\quad + (u^c \nabla_c V_\theta) (u_a u_b - g_{ab}).
\end{align}

Note that terms in the action 
linear in $\theta$ do not contribute to 
$V - \theta V_\theta$, hence they contribute 
only to the part of the stress tensor proportional
to the spatial metric $u_a u_b - g_{ab}$.
This can be traced to the contributions from the Lagrange
multiplier term in the action, but it can also be understood
more directly 
 from the fact that 
the metric enters such terms via the combination
$\sqrt{-g}\theta=\partial_a(\sqrt{-g}u^a)$.
In order to 
maintain the unit norm condition $g_{ab} u^a u^b=1$,  
a metric variation must be accompanied by  
an aether variation $\delta_\parallel u^a =-\tfrac12 (\delta g_{mn}\, u^mu^n) u^a$.
The combined $\delta g^{ab}$ and $\delta_\parallel u^a$ variations
yield
\begin{equation}
\delta(\sqrt{-g}u^a)=\tfrac12\sqrt{-g}u^a(-g_{mn}+u_mu_n)\delta g^{mn}.
\end{equation}
The corresponding contribution to the stress tensor is thus
proportional to the spatial metric, and corresponds to an isotropic pressure in the aether frame.

\section{Linearized perturbations and dispersion relations} \label{section:modes}

We consider linearized perturbations about the flat space-time solution with a constant aether field
\begin{equation} \label{flatbackground}
g_{ab} = \eta_{ab}, \quad u^a = \delta^a_0, \quad \varphi = 0.
\end{equation}
To ensure that (\ref{flatbackground}) is a solution of the equations of motion, the potential must satisfy
\begin{equation}
V(0,0) = V_\varphi(0,0) = 0.
\end{equation}
In order to study linear perturbations, the potential is expanded to second order in the fields
\begin{equation} \label{expandpotential}
V(\theta, \varphi) \approx \tfrac{1}{2} m^2 \varphi^2 + \mu M \theta \varphi,
\end{equation}
where $m$ is the mass of the scalar field. The constant $\mu$ has dimensions of mass, and we include a factor of $M$ to compensate for the canonical normalization of the aether perturbations.
The expansion (\ref{expandpotential}) does not include a term linear in $\theta$, because such a term is a total divergence and therefore does not affect the classical equations of motion. 
The term proportional to $\theta^2$ is also omitted, because it can be absorbed into the $c_2$ term in the Einstein-aether Lagrangian.

Because the background solution (\ref{flatbackground}) is invariant under spatial rotations, we can decompose the perturbations into irreducible representations of SO(3). The propagating degrees of freedom consist of a spin-2 graviton, a spin-1 aether-metric wave, a spin-0 aether-metric wave, and a spin-0 inflaton perturbation. At the linearized level, the presence of the scalar field affects only the spin-0 modes. This is because the term $\theta \varphi$ in the action 
depends only on the timelike and longitudinal parts of the aether field, both of which have spin-0.

We will adopt the variables and gauge choice of Foster \cite{Foster2006}. In this parametrization, the seven spin-0 degrees of freedom are labeled $h_{00}, \gamma, \phi, f, w^0, \nu, \varphi$, and a general spin-0 perturbation of the metric and aether can be written as
\begin{align}
g_{ab} &= \left[
\begin{array}{cc}
1 + h_{00} & \partial_i \gamma \\
\partial_i \gamma & -1 + \partial_i \partial_j \phi + \frac{1}{2}( \Delta f \delta_{ij} - \partial_i \partial_j f)
\end{array}
\right], \\
u^a &= [1 + w^0, \partial_i \nu].
\end{align}
Here $0$ labels the time coordinate,
latin indices $i,j, \ldots = 1,2,3$ are spatial, and $\Delta=\delta^{ij}\partial_i\partial_j$.
We can simplify the equations of motion by using the gauge freedom to set $\nu = \gamma = 0$  \cite{Foster2006}.
Additionally, we can eliminate $w^0$ using the constraint $u^a u_a = 1$, giving
\begin{equation}
w^0 = -\tfrac{1}{2} h_{00}.
\end{equation}
With these choices, and taking perturbations to have the space-time dependence $e^{i(kx - \omega t)}$ in terms of the flat space coordinates $(t,x,y,z)$, the metric and aether take the form
\begin{align}
g_{ab} &= \text{diag} \left(1+h_{00},-1- k^2 \phi,-1-\tfrac{1}{2} k^2 f,-1- \tfrac{1}{2} k^2 f \right), \\
u^a &= \left(1 - \tfrac{1}{2}h_{00}, 0, 0, 0 \right).
\end{align}

In terms of these fields the linearized versions of the scalar field equation, the (0,0) Einstein equation, the (1,1) Einstein equation and the sum of the (2,2) and (3,3) Einstein equations are
\begin{align} \label{eom1}
0 &= -\omega^2 \varphi +k^2 \varphi + m^2 \varphi - \tfrac{1}{2} k^2 i \omega \mu M (f + \phi), \\
0 &=  -k^2 f - c_{14} h_{00}, \\
0 &= (1 + c_2) k^2 \omega^2 f + c_{123} k^2 \omega^2 \phi + 2 \mu i \omega \varphi / M, \\
0 &= (1 + c_{13} + 2 c_2) k^2 \omega^2 f \nonumber \\
&   -k^2(k^2 f + 2 h_{00} - 2 (1 + c_2) \omega^2 \phi)+ 4 \mu i \omega \varphi / M. \label{eom4}
\end{align}
where we have adopted the notation $c_{14} = c_1 + c_4$, $c_{123} = c_1 + c_2 + c_3$, etc.

In the absence of the scalar field the linearized modes all satisfy a linear 
dispersion relation of the form $\omega = s_i k$ where $s_i$, 
$i = 0,1,2$ is a speed depending only on the spin of the mode. This is because the aether terms in the Lagrangian all contain two derivatives. The scalar field action contains terms with fewer than two derivatives, so in general 
the spin-0 modes
are dispersive. The dispersion relation is obtained by taking the determinant of Eqs.~\eqref{eom1}-\eqref{eom4} considered as a matrix equation. This can be simplified to
\begin{equation} \label{dispersion}
(\omega^2 - k^2 - m^2) (\omega^2 - s_0^2 k^2) + \frac{3 \kappa_C}{2} \mu^2 \omega^2 - \frac{s_0^2}{c_{123}} \mu^2 k^2 = 0,
\end{equation}
where $\kappa_C$ is given by (\ref{G_c}) and $s_0$ is the wave speed of the spin-0 mode when $\mu=0$ \cite{Jacobson2004}
\begin{equation} \label{s0}
s_0^2 = \frac{(2 - c_{14}) c_{123}}{(1 - c_{13}) (2 + c_{13} + 3 c_2) c_{14}}.
\end{equation}
The fact that there are just two, rather than four, propagating spin-0 modes is a result of diffeomorphism invariance,
and is reflected by the fact that Eq.~\eqref{dispersion} is quadratic in 
$\omega^2$ and therefore admits two solutions.
In general these modes are superpositions of the spin-0 parts of the metric, aether and scalar fields.
Their polarizations can be obtained by solving Eqs.~\eqref{eom1}-\eqref{eom4} with fixed $\omega$ and $k$; they depend on the wavenumber and are valid only in our chosen gauge.

In the homogeneous limit $k \to 0$ the dispersion relation has the form
\begin{equation}
\omega^2(\omega^2 - m^2 + \tfrac{3}{2} \kappa_C \mu^2) = 0,
\end{equation}
which shows that one mode remains gapless, while the mass of the other mode is modified by the $\mu$ term.

\section{Linearized stability and Cherenkov constraint} \label{section:stability}

In what follows we will consider stability of the classical theory linearized about flat space-time. Although this is not sufficient to guarantee stability of the full interacting quantum theory, or even of the linearized theory about a curved background, it is a natural physical condition to impose and will lead to constraints on the parameters $c_{1,2,3,4}, m, M, \mu$. In the following three sections we will show that the linearized theory is stable provided $\mu$ is not too large compared to $m$, and the aether parameters lie within the allowed parameter range for Einstein-aether theory.

To say a theory is stable is to say that regular ``initial data'' does not grow exponentially in ``time''. This notion presumes that the theory admits an initial value formulation with respect to some foliation of the space-time by surfaces, called Cauchy surfaces. If the theory is stable with respect to one fiducial such foliation, then that suffices to establish stability from a physical point of view. This is because regular initial data on any other Cauchy surface can be evolved back to the fiducial Cauchy surface, where it will again be regular, hence the stability of its complete evolution is assured.

In the case of linearized Einstein-aether theory, the field equations are of hyperbolic type, but not with respect to the space-time metric, since different modes travel at different speeds. Nevertheless there is a maximum speed for the modes, and this determines the class of time functions whose constant-time surfaces are Cauchy surfaces for the full set of equations of motion. In particular, the surfaces orthogonal to the aether vector in a Minkowski background can serve as Cauchy
surfaces for the purpose of establishing stability.

We should note that Carroll et. al.  \cite{Carroll2009} have proposed a different criterion for stability, namely that linearized perturbations that are oscillatory on any Lorentz time slice should not grow with respect to the corresponding time coordinate, and they showed that this criterion is strong enough to rule out Einstein-aether theory for most values of the parameters $c_{1,2,3,4}$. 
We believe that this criterion is not the physically relevant one, because it corresponds to imposing stability with respect to a time function whose constant-time surfaces are not Cauchy surfaces for the equations.
This issue will be discussed more fully in a forthcoming 
publication \cite{Donnelly2010}.

We therefore impose stability in the aether frame. This means that the dispersion relation (\ref{dispersion}) must have two real solutions $\omega_{\pm}$ for every real wavenumber $k$. Equation (\ref{dispersion}) is a quadratic equation in $\omega^2$ with real coefficients; 
its roots $\omega_\pm^2$ are real 
provided the discriminant is positive. This is necessary but not sufficient, since $\omega_\pm^2$ must be positive in order for $\omega_\pm$ to be real. We will use the fact that two real numbers $\omega_\pm^2$ are both positive if and only if the sum $\omega_-^2 + \omega_+^2$ and product $\omega_+^2 \omega_-^2$ are both positive. This will be convenient to impose since the sum and product of roots can be read directly from the coefficients of the dispersion relation. In what follows we will work backward, first requiring the sum and product roots to be positive and then returning to require the discriminant to be positive.

The sum and product of the roots are given by
\begin{align} \label{rootsum}
\omega_+^2 + \omega_-^2 &= (1 + s_0^2) k^2 + \left(m^2 - \tfrac{3}{2} \kappa_C \mu^2 \right), \\
\label{rootproduct}
\omega_+^2 \omega_-^2 &= s_0^2 k^4  + (m^2 - \mu^2 / c_{123}) s_0^2 k^2.
\end{align}
These will be non-negative for all $k$ if and only if
\begin{align}
s_0^2 &\geq 0, \label{constraint0} \\
\kappa_C \mu^2 &\leq 2 m^2 / 3, \label{constraint1} \\
\mu^2 / c_{123} &\leq m^2. \label{constraint2} 
\end{align}

Finally, we impose the condition that the roots $\omega_{\pm}^2$ of the quadratic (\ref{dispersion}) are both real. Thus we require that the discriminant be positive,
\begin{align} \label{disc}
&(1 - s_0^2)^2 k^4 + 2 [(1 + s_0^2)(m^2 - \tfrac{3}{2} \kappa_C \mu^2) \nonumber \\
	& {} - 2 (m^2 -  \mu^2 / c_{123}) s_0^2] k^2 + (m^2 - \tfrac{3}{2} \kappa_C \mu^2)^2 \geq 0.
\end{align}
The discriminant (\ref{disc}) is of the form $a k^4 + b k^2 + c$ with $a, c \geq 0$. It will be positive if $b \geq 0$ or $b^2 - 4 a c \leq 0$, in other words if $b \geq - 2 \sqrt{ac}$. This condition is
\begin{equation}
(1 + s_0^2 + \abs{1 - s_0^2})(m^2 - \tfrac{3}{2} \kappa_C \mu^2) - 2 (m^2 - \mu^2 / c_{123}) s_0^2 \geq 0,
\end{equation}
where we have made use of (\ref{constraint1}) to simplify the expression.
This inequality can be simplified into two cases depending on the sign of $1 - s_0^2$:	
\begin{align} \label{constraint3a}
1 / c_{123} &\geq 3 \kappa_C / 2 &\text{if } s_0 \geq 1, \\	
\label{constraint3b}
m^2 (1 - s_0^2) &\geq (3 \kappa_C/2 - s_0^2/c_{123}) \mu^2 &\text{if } s_0 < 1.
\end{align}

We now consider a further physical constraint on the theory: If the phase velocity of scalar aether waves is less than the speed of light, then highly energetic particles can Cherenkov radiate in vacuum. Observations of cosmic rays strongly constrain this behavior \cite{Elliott2005}. We therefore require the phase velocity of aether waves to be $\geq 1$ for all $k$.

First we show that if a mode has phase velocity $\geq 1$ for some $k$ then its phase velocity remains $\geq 1$ for all $k$. If a mode becomes subluminal at a particular wavenumber $k$ it would have to satisfy the dispersion relation with $\omega^2 = k^2$,
\begin{equation}
[m^2(s_0^2 - 1) + (\tfrac{3}{2} \kappa_C - s_0^2 / c_{123})\mu^2] k^2 = 0.
\end{equation}
If this equation holds for any $k > 0$ then it holds for all $k$. This means that if a given mode has speed $\geq 1$ for any $k$, it has speed $\geq 1$ for all $k$.

It is therefore sufficient to enforce that the modes are superluminal in the limit $k \to \infty$. To find the phase velocities in this limit, we express (\ref{dispersion}) as a polynomial in $\omega$ and keep only the dominant power of $k$ in each coefficient. The resulting mode speeds are simply $1$ and $s_0$, so the Cherenkov constraint is simply $s_0 \geq 1$, as it is in pure Einstein-aether theory.

Thus when the Cherenkov constraint is imposed, (\ref{constraint3a}) is the condition for stability. 
Equations (\ref{constraint1}), (\ref{constraint2}) and (\ref{constraint3a}) together with $s_0^2 \geq 1$ are therefore necessary and sufficient conditions for stability and absence of vacuum Cherenkov radiation.

We now derive the condition for the group velocity 
of the linearized waves to be positive. 
This will play a role in the energy positivity constraint. 
In order for the waves to have positive group velocity, 
we must have $d \omega^2 / d k^2 > 0$. 
In the limit $k \to \infty$, the group velocity is positive. 
Since $d \omega^2 / d k^2$ is a continuous function 
of $k$, it is sufficient to show that there is no solution 
of the dispersion relation for which $d \omega^2 / d k^2 = 0$.

Solving $d \omega^2 / d k^2 = 0$ yields
\begin{equation}
2 s_0^2 k^2 = ( \mu^2 / c_{123} - m^2) s_0^2 + (1 + s_0^2) \omega^2.
\end{equation}
Solving for $k$, and substituting back into the 
dispersion relation gives $a \omega^4 + b \omega^2 + c = 0$ where
\begin{align}
a &= (s_0^2 - 1)^2, \\
b &= 2 s_0^2 [(1 - s_0^2) m^2 + (1 + s_0^2) \mu^2 / c_{123} - 3 \kappa_C \mu^2], \\
c &= (m^2 - \mu^2/c_{123})^2 s_0^2.
\end{align}
Assuming (\ref{constraint2}) and $s_0^2 \geq 0$ implies $c \geq 0$. 
Just as in the discussion following (\ref{disc}), 
there are no real roots for $\omega$ (i.e. no positive roots for $\omega^2$) if and only if $b > -2 \sqrt{a c}$, where
\begin{equation}
b + 2 \sqrt{a c} = 4 s_0^2 (1 / c_{123} - 3 \kappa_C / 2)\mu^2.
\end{equation}
Therefore the group velocity is positive for all $k$ if and only if
Eqs. (\ref{constraint0}) and (\ref{constraint3a}) hold, as required by stability. 

\section{Positivity of energy density} \label{section:energy}

In addition to real frequencies, we further require that the linearized perturbations have positive energy. 
In doing so we necessarily run into the issue that there is no suitable local covariant expression for the energy density of a diffeomorphism-invariant theory that accounts for the energy in the gravitational field.
However for wave-like linear perturbations that are periodic in time, it is possible to define the average energy density.
The positive energy conditions for Einstein-aether theory have been found both by pseudotensor methods \cite{Eling2005} and using the Noether current \cite{Foster2006}. The two methods can be shown to produce equivalent results \cite{Iyer1994}; we will follow the latter approach, which is simpler both conceptually and computationally.

The Noether current is defined as follows \cite{Iyer1994}.
We first define the canonical one-form $\theta^a$ 
via the total divergence that arises when varying the action
\begin{align}
\delta S = \int \sqrt{\abs{g}} d^4x \left( E[\psi] \cdot \delta \psi + \nabla_a \theta^a[\delta \psi] \right),
\end{align}
where $E$ is the equation of motion, and $\psi$ schematically denotes collectively all the dynamical fields. 
The Noether current 1-form $J^a[\xi]$ associated to the vector field $\xi^a$ is then given by
\begin{equation} \label{current}
J^a[\xi] = \theta^a[\mathcal{L}_\xi \psi] - L \xi^a,
\end{equation}
where $\mathcal{L}_\xi$ is the Lie derivative, and $L$ is the Lagrangian.
When the field equation $E = 0$ is satisfied the Noether current $J^a$ constructed this way is conserved,
\begin{equation}
\nabla_a J^a = 0.
\end{equation}
The energy is the Noether charge associated with the asymptotic time translation $t^a$. It is obtained by choosing $\xi^a$ to coincide with $t^a$ at infinity and integrating the flux of the corresponding current $J^a$ through a Cauchy surface $\Sigma$,
\begin{equation}
\mathcal{E} = \int_\Sigma J^a d\Sigma_a,
\end{equation}
where $d\Sigma_a$ is the induced volume form on $\Sigma$.

The Noether current is a sum of two terms $J_{\text \ae}$ and $J_\varphi$ which correspond to contributions from $L_{\text \ae}$ and $L_\varphi$ respectively. 
The Noether current $J_{\text \ae}$ was found by Foster \cite{Foster2005}. 
To evaluate $J_\varphi$ it is useful to integrate the linearized Lagrangian by parts, giving a new Lagrangian
\begin{equation}
L_\varphi' = \tfrac{1}{2} \nabla_a \varphi \nabla^a \varphi - \tfrac{1}{2} m^2 \varphi^2 + \mu M u^a \nabla_a \varphi.
\end{equation}
With suitable asymptotic boundary conditions this Lagrangian leads to the same equations of motion. It also gives the same Noether charge, and therefore the same time-averaged energy density, even though it does not give the same Noether current.
The advantage of $L_\varphi'$ over $L_\varphi$ is that it contains no derivatives of $u$ or $g$, so the corresponding canonical one-form contains only terms proportional to $\delta\varphi$, not to $\delta u$ or $\delta g$.
Varying the action, we find the contribution of $L_\varphi'$ to the canonical one-form is
\begin{equation}
\theta^a_\varphi[\delta \varphi] = (\nabla^a \varphi + \mu M u^a) \delta \varphi.
\end{equation}
The corresponding Noether current is then determined from (\ref{current}),
\begin{equation} \label{current_phi}
J^a_\varphi[\xi]= (\nabla^a \varphi + \mu M u^a) \dot \varphi - L_\varphi' \xi^a.
\end{equation}

We use the Noether current to find the energy density of linearized waves following Foster \cite{Foster2006}. Consider a compact source in an asymptotically flat space-time. Fixing $R$ to be a sphere of large coordinate radius $r$, the rate at which energy is radiated from the source is given by the flux of the Noether charge
\begin{equation}
-\mathcal{\dot E} = \int_R \vec{J} \cdot d\vec{A},
\end{equation}
where $\vec{J}$ is the spatial part of $J^a$. The energy is being carried away by waves with average energy density $u$ and group velocity $v_g$ (which we assume positive), so that the average rate of energy loss is
\begin{equation} 
- \avg{\mathcal{\dot E}} = \int_R u \, v_g \, dA,
\end{equation}
where $\avg{\cdot}$ denotes time averaging over one period of the wave. Equating the two expressions for the rate of energy loss we find
\begin{equation}
u = \avg{J_r} / v_g.
\end{equation}
To find the total energy density we need to evaluate this formula with the total Noether current, $J_\varphi + J_{\text \ae}$.

To carry out this calculation, note that for sufficiently large $r$ all of the dynamical fields can be approximated by spherical waves for which
\begin{equation} \label{spherical_wave}
\partial_i \psi = -(1 / v_p) \dot \psi \hat{r}_i,
\end{equation}
where $v_p$ is the phase velocity of the mode and $\hat{r}$ is the outward facing unit normal.
Equation (\ref{spherical_wave}) allows spatial derivatives in the energy density to be exchanged for time derivatives. The contribution to the energy density coming from $J_\varphi$ is
\begin{equation}
u_\varphi = \avg{\dot \varphi ^2} / (v_p v_g).
\end{equation}
To find the total energy density we add this to the contribution to the energy density from the Lagrangian $L_{\text \ae}$ \cite{Foster2006}. Up to a positive factor we obtain
\begin{equation}\label{u0}
u \propto M^2 (2 - c_{14}) k^4 \avg{\dot f^2} + 8 c_{14} \avg{\dot \varphi^2}.
\end{equation}	
It follows that $0 \leq c_{14} \leq 2$ is a sufficient condition for positive energy density. We cannot yet conclude that it is a necessary condition, since $\varphi$ and $f$ are related by the linearized equations of motion (\ref{eom1}-\ref{eom4}), which imply
for the complex amplitudes
\begin{equation} 
\left[c_{123} (\omega^2 - k^2 - m^2) + \mu^2 \right] \varphi = \tfrac{1}{2} \mu M (1 - c_{13}) i \omega k^2 f.
\end{equation}
Substituting this into
(\ref{u0})
we find that up to a positive multiplicative factor the energy density is given by
\begin{equation} \label{ufinal}
(2 - c_{14}) \left[ c_{123} (\omega^2 - k^2 - m ^2) + \mu^2 \right]^2 + 2 c_{14} \mu^2 (1 - c_{13})^2\omega^2,
\end{equation}
where $\omega$ and $k$ are related by the dispersion relation (\ref{dispersion}). This expression holds for both of the spin-0 modes, each of which corresponds to a different solution of the dispersion relation.

We can now consider two separate limits of the dispersion relation. In the homogeneous limit $\omega = k = 0$, the second term in (\ref{ufinal}) drops out, and positivity requires $c_{14} \leq 2$. In the limit $k \to \infty$ with $\omega = k$, the first term is sub-leading in $k$, so that $c_{14} \geq 0$ is also needed.
The upshot of this calculation is that the constraint from energy positivity of the spin-0 modes is
\begin{equation}
0 \leq c_\text{14} \leq 2,
\end{equation}
which is 
the same as in Einstein-aether theory 
without the scalar coupling.

\section{Combined constraints} \label{section:constraints}

In this section we first gather together all the 
constraints derived from stability, Cherenkov radiation, 
and energy positivity.
Next we consider the constraints from post-Newtonian effects,
and finally combine these with the other constraints to
determine the allowed parameter region.

\subsection{Stability, Cherenkov and energy constraints}

The constraints discussed already, together with 
those for the spin-1 and spin-2 modes \cite{Jacobson2008}, 
which are not affected 
by the presence of the scalar field, are as follows:
\begin{align}
\text{spin-0 stability} \quad & \frac{3\kappa_C}{2}\le\frac{1}{c_{123}}\le\frac{m^2}{\mu^2},  \label{spin0stab} \\
\text{spin-0 speed} \quad &\frac{(2 - c_{14}) c_{123}}{(1 - c_{13}) (2 + c_{13} + 3 c_2) c_{14}} \geq 1,  \label{spin0speed} \\
\text{spin-1 speed} \quad & \frac{2c_1 - c_1^2 + c_3^2} {2 c_{14} (1 - c_{13})} \geq 1, \label{spin1speed} \\
\text{spin-2 speed} \quad & \frac{1}{1 - c_{13}} \geq 1, \label{spin2speed} \\
\text{spin-0 energy} \quad & 0 \leq c_{14} \leq 2, \label{spin0energy} \\
\text{spin-1 energy} \quad & (2 c_1 - c_1^2 + c_3^2)(1 - c_{13}) \geq 0, \label{spin1energy} \\
\text{spin-2 energy} \quad & M^2 \geq 0. \label{spin2energy}
\end{align}
Since the spin-1 and spin-2 modes satisfy a dispersion relation of the form $\omega^2 = s^2 k^2$, the stability constraint for each of these modes is implied by the Cherenkov constraint $s^2 \geq 1$. By contrast, the spin-0 modes have a $k$ dependent
phase velocity, so the separate stability
and Cherenkov constraints involve several conditions.

The first inequality in (\ref{spin0stab}) is 
actually implied by
the other inequalities. To see this, note that (\ref{spin0speed}), 
(\ref{spin2speed}), and (\ref{spin0energy}) imply that 
$c_{123}$ and $\kappa_C$ (\ref{G_c}) must have the same sign.
The first inequality in (\ref{spin0stab}) therefore 
reduces to the inequality $c_{13}\le1$, which is already
implied by (\ref{spin2speed}). 

\subsection{Post-Newtonian parameters} \label{section:ppn}

As we mentioned in Sec.~\ref{section:introduction}, Einstein-aether theory has a Newtonian limit with a renormalized gravitational constant $G_N$ given in (\ref{G_n}). The spin-2 energy constraint says that $G$ must be positive. Combined with the spin-0 energy constraint, this also implies that $\kappa_N$ and therefore 
Newton's constant $G_N$ (\ref{G_n}) is also positive. 

The
post-Newtonian parameters $\beta$ and $\gamma$ 
take the same values (unity) as in general relativity \cite{Eling2003}. 
The effects of Lorentz violation are captured at first post-Newtonian order by the dimensionless preferred frame parameters $\alpha_1$ and $\alpha_2$.
These can be expressed in terms of $c_{1,2,3,4}$ as \cite{Graesser2005,Foster2005b}
\begin{eqnarray}
\alpha_1 &=& \frac{-8 (c_3^2 + c_1 c_4)} {2 c_1 - c_1^2 + c_3^2}, \label{alpha1} \\
\alpha_2 &=& \frac{\alpha_1}{2} - \frac{(c_1 + 2 c_3 - c_4)(2 c_1 + 3 c_2 + c_{34})}{c_{123}(2 - c_{14})}. \label{alpha2}
\end{eqnarray}
 The preferred frame parameters are constrained observationally to be small  
 \cite{Will2005}, 
 $\alpha_1\lesssim 10^{-4}$ and $\alpha_2\lesssim 4\times10^{-7}$,
 and because Einstein-aether theory has four free parameters they can be made to vanish exactly in a two-dimensional subspace of parameter space. We will see that similar considerations apply also in the presence of the scalar field.

We now consider the effect on the post-Newtonian expansion of coupling to the scalar field $\varphi$. 
The stability constraint (\ref{spin0stab}) shows that $\varphi$ cannot be massless if the cosmological Newton constant is to be positive. 
In order to simplify our analysis, we will further assume that $m$ is large enough so that the Compton wavelength of 
$\varphi$ particles is much smaller than 
any scale on which gravity has been tested in the late universe.
Newton's law of gravitation has been tested to sub-millimeter scales, 
so we shall assume the lower bound $m \gtrsim 10^{-3} \text{eV}$
on the mass. If the mass is smaller than this, a more complete 
analysis would be required to determine the effects of the scalar field
coupling to gravity. Most likely the mass of the inflaton must
be tremendously larger than this anyway. 

Assuming that all fields vary on scales much larger than $1/m$, we can integrate out the scalar field. We first solve the linearized scalar field equation (\ref{klein-gordon}) by an expansion in $\square / m^2$,
\begin{equation} \label{derivative_expansion}
\varphi = - (\square + m^2)^{-1} \mu M \theta = - \frac{\mu M}{m^2} \theta + \frac{\mu M}{m^4} \square \theta + \ldots,
\end{equation}
where in what follows we will keep only the leading order contribution, $\varphi = -(\mu M/m^2) \theta$. If we substitute this value back into the linearized scalar field action and expand to $O(\varphi^2)$ we find
\begin{align}
L_\varphi 
&= \frac{\mu^2 M^2}{2 m^4} \nabla_a \theta \nabla^a \theta + \frac{\mu^2 M^2}{2 m^2} \theta^2.
\end{align}
The $(\nabla \theta)^2$ term is of higher order in the derivative expansion, and so will be neglected. The remaining term can be absorbed in the $c_2$ term of the aether action, which is given by $-\frac{1}{2} M^2 c_2 \theta^2$. Therefore, upon integrating out the scalar field, the new Lagrangian is simply $L_{\text \ae}$ with a new value of the parameter $c_2$ which we denote $c_2'$,
\begin{equation}
c_2 \to c_2' = c_2 - \frac{\mu^2}{m^2}. \label{c2shift}
\end{equation}
The higher derivative terms in the expansion (\ref{derivative_expansion}) lead to higher derivative terms in the aether action. If we neglect these higher-order corrections, the post-Newtonian parameters can be determined from the known results for Einstein-aether theory \eqref{alpha1} and \eqref{alpha2} using $c_2'$ in place of $c_2$.

Just as in pure Einstein-aether theory, 
we can set both preferred frame parameters $\alpha_1, \alpha_2$ to zero,
now by choosing 
\begin{align}
c_2 &= \frac{-2c_1^2 -c_1 c_3 + c_3^2}{3 c_1} + \frac{\mu^2}{m^2}, \label{c2} \\
c_4 &= -\frac{c_3^2}{c_1}. \label{c4}
\end{align}
Once this choice of $c_4$ has been made, the 
spin-1 and spin-2 speed constraints and the 
spin-0 and spin-1 energy constraints \eqref{spin1speed}-\eqref{spin1energy}
can be conveniently expressed in terms of $c_\pm \equiv c_1 \pm c_3$, 
and they hold if and only if $0 \leq c_+ \leq 1$ and $c_- \geq 0$
\cite{Foster2005b}. Moreover, once this choice of 
$c_2$ has been made, we have
\beq\label{c123}
c_{123}=\frac{c_{13}^2}{3c_1}+\frac{\mu^2}{m^2},
\eeq
which is positive when the other constraints hold,
since the positivity of $c_1$ is implied by 
$c_{\pm}\ge0$. 
The second inequality of (\ref{spin0stab}) then becomes
\begin{equation} \label{mu_over_mu_max}
\mu^2 \leq c_{123} m^2 = \mu^2 + \frac{c_{13}^2}{3 c_1} m^2,
\end{equation}
which is also automatically satisfied when the other constraints
are imposed. 

\begin{figure}[t]
\begin{tabular}[t]{cc}
\raisebox{3.75cm}{$c_-$} & \includegraphics[width=7.5cm]{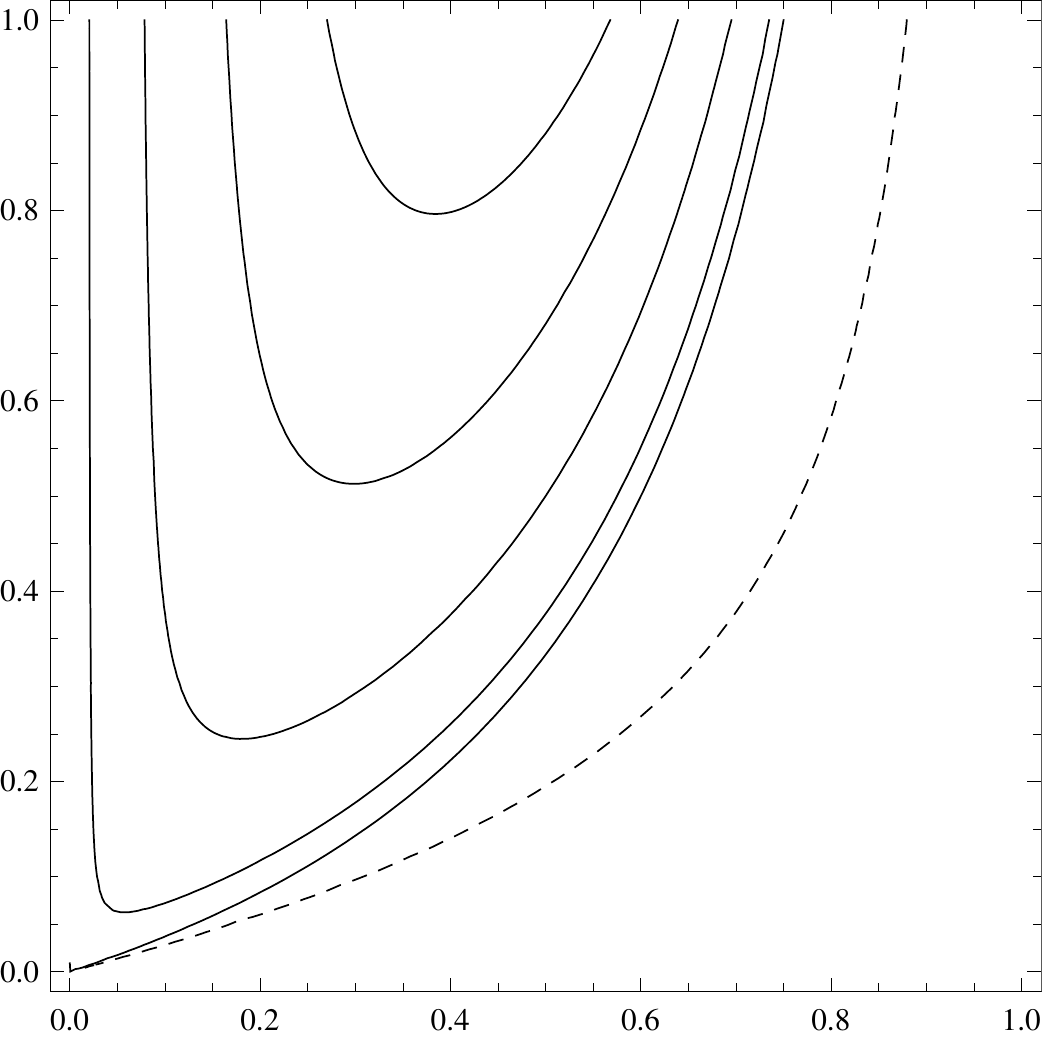} \\
& $\qquad c_+$
\end{tabular}
\caption{ \label{parameter_range}
The allowed ranges for the parameters $c_\pm=c_1\pm c_3$, 
assuming $c_2$ and $c_4$ are chosen so that the preferred
frame PPN parameters vanish. The spin-0 Cherenkov constraint
is satisfied below the solid lines, which correspond (from lowest to highest)
to $\mu = 0$, $\mu = \frac{1}{4}m$, $\mu = \frac{1}{2}m$, $\mu = m$, and $\mu/m \to \infty$. The dashed line corresponds to the parameter values for which the damping rate from weak-field binary systems calculated in \cite{Foster2006} agrees with general relativity.
}
\end{figure}

The condition for absence of vacuum Cherenkov radiation into the spin-0 mode (\ref{spin0speed}) depends on $c_2$, and therefore on 
$\mu$ via (\ref{c2}). 
In pure Einstein-aether theory (i.e. when $\mu = 0$), the value of $c_2$ in Eq.~\eqref{c2} goes to zero as $c_\pm \to 0$. 
In this limit, $s_0$ (\ref{s0}) depends on the ratio between $c_+$ and $c_-$. 
However, if $\mu \neq 0$, then $c_2$ does not go to zero, and $s_0 \to \infty$ as $c_\pm \to 0$. This means that the Cherenkov constraint can always be satisfied in an open neighborhood of $c_\pm = 0$. 
As shown in Fig.~\ref{parameter_range}, the range of allowed $c_\pm$ is significantly enlarged as $\mu$ increases.

\subsection{Other constraints}

Further constraints arise from the effective cosmological
value of Newton's constant in the Friedmann equation, 
and radiation and orbital dynamics of compact binaries.

As explained above, when the PPN parameters
$\alpha_{1,2}$ have been set to zero by 
the choices \eqref{c2} and \eqref{c4}, it follows from the
other constraints that 
$c_{123}\ge0$.
Then (\ref{spin0speed}) also implies that
$\kappa_C$ (\ref{G_c}) is positive. This would be 
the value of the cosmological Newton constant
relevant during inflation. For later times, after the
inflaton is integrated out\footnote{The derivative terms
in the Klein-Gordon equation are
presumably negligible when $H \ll m$, which occurs at a 
temperature $T \ll \sqrt{mM}$. This is satisfied at the
nucleosynthesis temperature $\sim 0.1$ MeV as long as
$m \gg 10^{-18}$ eV.}, the
relevant value is $G'_C=\kappa'_C G$,
where $\kappa_C'$ is given by (\ref{G_c}) but with $c_2'$ in place of 
$c_2$.  
When $c_2$ and $c_4$ are set according to (\ref{c2},\ref{c4}) 
it follows \cite{Foster2005b} that $\kappa'_C = \kappa_N$. 
Predictions of primordial element abundances from big bang nucleosynthesis are sensitive to $G'_C$, and 
agreement with observations requires \cite{Carroll2004}
that $G'_C$ not differ too much from earthbound measurements
of Newton's constant,
\begin{equation} \label{bbn}
\abs{G'_C / G_N - 1} \lesssim 1/8.
\end{equation}
This constraint is thus trivially satisfied when the 
preferred frame PPN parameters vanish. 

The rate of change in orbital period of binary pulsars due to gravitational and aether radiation, ignoring effects from strong self-gravity of the bodies, was computed by Foster \cite{Foster2006}.
The observed orbital decay agrees with that found in general relativity to within better than one percent \cite{Will2005}.
The constraint that the rate in weak field Einstein-aether theory agree with that found in general relativity is indicated in Fig.~\ref{parameter_range}. 
The strong-field effects that contribute to the damping rate and orbital dynamics of binary systems was computed in \cite{Foster2007} in terms of as yet unknown parameters that characterize the velocity dependence of the energy (or action) of the bodies.
Strong-field effects are likely to lead to further constraints on the parameters, restricting $c_\pm$ to be less than $\sim 0.01-0.1$.

\section{Homogeneous cosmology} \label{section:cosmology}

We consider now the cosmology of Einstein-aether theory
coupled to a single scalar inflaton field, with a potential
that depends on the expansion of the aether.
The metric is assumed to be homogeneous,
isotropic, and spatially flat, hence it can be written in the form
\begin{equation} \label{frw}
ds^2 = dt^2 - a(t)^2 d\vec{x}^2.
\end{equation} 
In a homogeneous and isotropic solution the aether must be aligned with the cosmological rest frame.
In this case, the spatial components of the aether equation are automatically satisfied,
and the time component of the aether equation of motion (\ref{u}) just 
determines the Lagrange multiplier $\lambda$.
That the aether relaxes to such an isotropic configuration during cosmological expansion
is shown perturbatively in \cite{Lim2004,Kanno2006}. A nonlinear extension of this analysis is currently underway \cite{Carruthers2010}.

We note that our results for homogeneous cosmology
apply as well to the extended version of Ho\v{r}ava gravity \cite{Horava2009}
proposed in Ref.~\cite{Blas2009}. As shown in Ref.~\cite{Jacobson2010}, this follows
because the aether in these solutions is hypersurface-orthogonal.

With the metric (\ref{frw}), the expansion is $\theta = 3H$ 
and the homogeneous scalar (inflaton) field equation (\ref{klein-gordon}) is
\begin{equation} \label{frw-scalar}
\ddot\varphi + \theta \dot\varphi + V_\varphi = 0,
\end{equation}
where the driving force $-V_\varphi$ is now dependent on the expansion $\theta$. 
The terms in $V_\varphi$ containing positive powers of $\theta$ lead to driving 
forces that are most relevant during inflation when the expansion is large. 
As the expansion slows, these forces subside, allowing for a graceful end to inflation.

The Friedmann equation can be derived by considering a metric of the form $ds^2 = N(t)^2 dt^2 - a(t)^2 d\vec{x}^2$ and varying the action with respect to $N$.
The normalization condition for $u$ together with the symmetry completely fixes $u$ to $u^a = (1/N, 0, 0 ,0)$, and the symmetry reduced Lagrangian density is (up to a total derivative)
\begin{equation} \label{reduced_lagrangian}
\sqrt{-g}L = N a^3 \left( -\frac{M_C^2}{3} \theta^2 + \frac{1}{2N^2} \dot \varphi^2 - V(\theta, \varphi) \right),
\end{equation}
where $\theta = 3H / N$ and we have introduced a ``reduced cosmological Planck mass'' $M_C \equiv M / \sqrt{\kappa_C}$ with $\kappa_C$ defined in (\ref{G_c}).
For terms in the action that are linear in $\theta$, the factor of $1/N$ in $\theta$ cancels the factor $N$ in the determinant of the metric, so that such terms do not contribute to the Friedmann equation. This is in agreement with the general form of the stress tensor (\ref{scalar-stress}).

Varying (\ref{reduced_lagrangian}) with respect to $N$ gives the Friedmann equation
\begin{equation} \label{friedmann}
\theta^2 = \frac{3}{M_C^2} \left( \tfrac{1}{2}\dot\varphi^2 + V - \theta V_\theta \right),
\end{equation}
where the gauge condition $N=1$ has been adopted.
This equation determines $\theta$ only 
implicitly as a function of $\varphi$ and $\dot\varphi$, since the potential $V$ also depends on $\theta$.

We now focus on the case when the potential has only quadratic terms,
\begin{equation}
V(\theta, \varphi) = \tfrac{1}{2} m^2 \varphi^2 + \mu M \theta \varphi,
\end{equation}
with $\mu>0$.
The Friedmann equation (\ref{friedmann}) then
has the standard form for a massive scalar.
While the $\theta \varphi$ 
coupling does not show up in the Friedmann equation, it
does affect the Einstein equation, via the pressure term in the 
scalar stress tensor (\ref{scalar-stress}), 
\beq
p=\dot{V_\theta}= \mu M \dot{\varphi}.
\eeq
The inflaton field equation (\ref{frw-scalar}) becomes
\begin{equation} \label{frw-scalar-2}
\ddot\varphi + \theta \dot\varphi + m^2\varphi +\mu M\theta = 0.
\end{equation}
The last term acts as an external force
that pushes $\varphi$ in the negative direction.

Slow roll solutions can be obtained by neglecting the kinetic term 
$\dot \varphi^2$ in (\ref{friedmann}), and the term $\ddot \varphi$ 
in the Klein-Gordon equation (\ref{frw-scalar}). 
The slow roll equations in this case become
\begin{eqnarray}
\theta&=&\sqrt{\frac32}\frac{m}{M_C}|\varphi|,\label{slowroll-theta}\\
\dot\varphi 
&=& - \sqrt{ \frac{2}{3}} m M_C \left(  \text{sgn}(\varphi) +\frac{\mu}{\mu_c}  \right), \label{slowroll-varphi}
\end{eqnarray}
where we have defined
\begin{equation} \label{mu_c}
\mu_c = \sqrt{\frac{2}{3 \kappa_C}} m = \sqrt{\frac{2 + c_{13} + 3 c_2}{3}} m.
\end{equation}

This slow roll solution need not be stable: If the Hubble force $-\mu M\theta$
is dominant then there is a feedback effect, where this 
force leads to larger field values, which leads to a more 
rapid expansion and therefore a stronger Hubble force.
The slow roll solution will be stable provided that 
$\varphi$ and $\dot\varphi$ are of the opposite sign for all 
values of $\varphi$, i.e.\ if $\abs{\mu} \leq \mu_c$. 
This condition is implied by the condition for stability 
of the linearized modes (\ref{spin0stab}).
This is not surprising, since that condition was originally inferred from 
the zero wavevector limit of the dispersion relation (\ref{constraint1}),
and we have restricted even in the nonlinear analysis to a quadratic potential.  

\begin{figure*} 

\subfloat[$\mu = 0$]{\label{phase1}
\begin{tabular}{cc}
\raisebox{3.5cm}{ $\frac{\displaystyle \dot\varphi}{m M_C}$ } & 
\includegraphics[width=7cm]{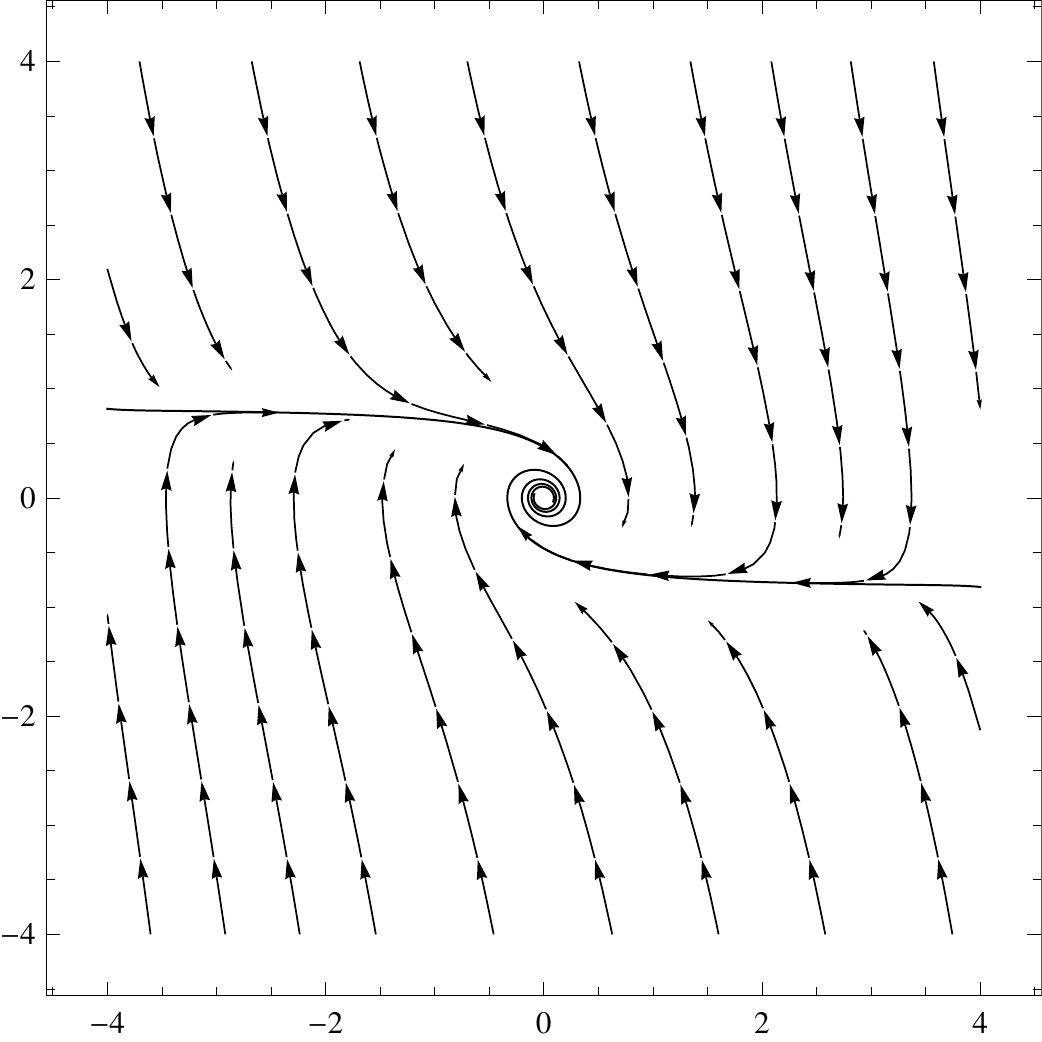} \\
& $\quad \frac{\displaystyle \varphi}{M_C}$ \\
\end{tabular}
}
\qquad
\subfloat[$\mu = \frac{1}{2} \mu_c$]{\label{phase2}
\begin{tabular}{cc}
\raisebox{3.5cm}{ $\frac{\displaystyle \dot\varphi}{m M_C}$ } & 
\includegraphics[width=7cm]{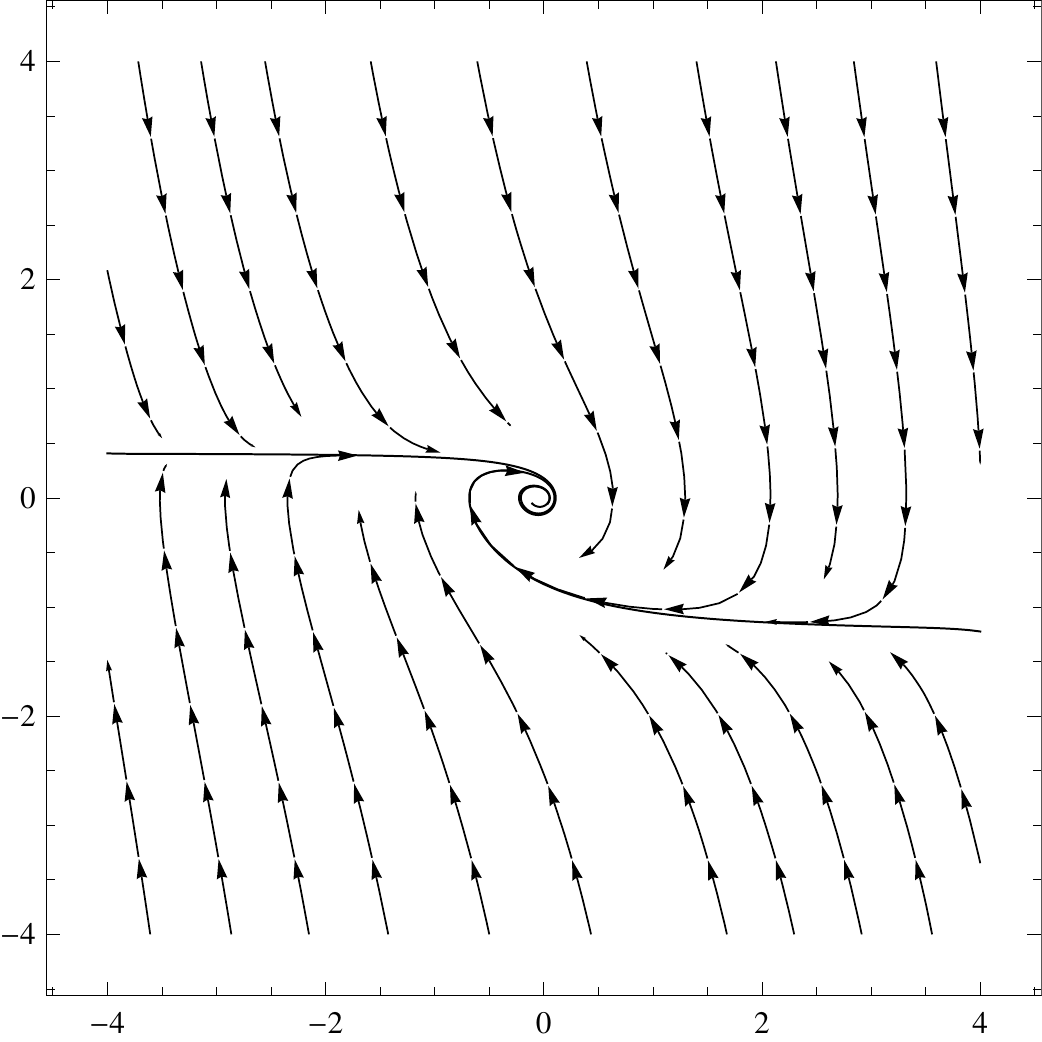} \\
& $\quad \frac{\displaystyle \varphi}{M_C}$ \\
\end{tabular}
}
\\
\subfloat[$\mu = \mu_c$]{\label{phase3}
\begin{tabular}{cc}
\raisebox{3.5cm}{ $\frac{\displaystyle \dot\varphi}{m M_C}$ } & 
\includegraphics[width=7cm]{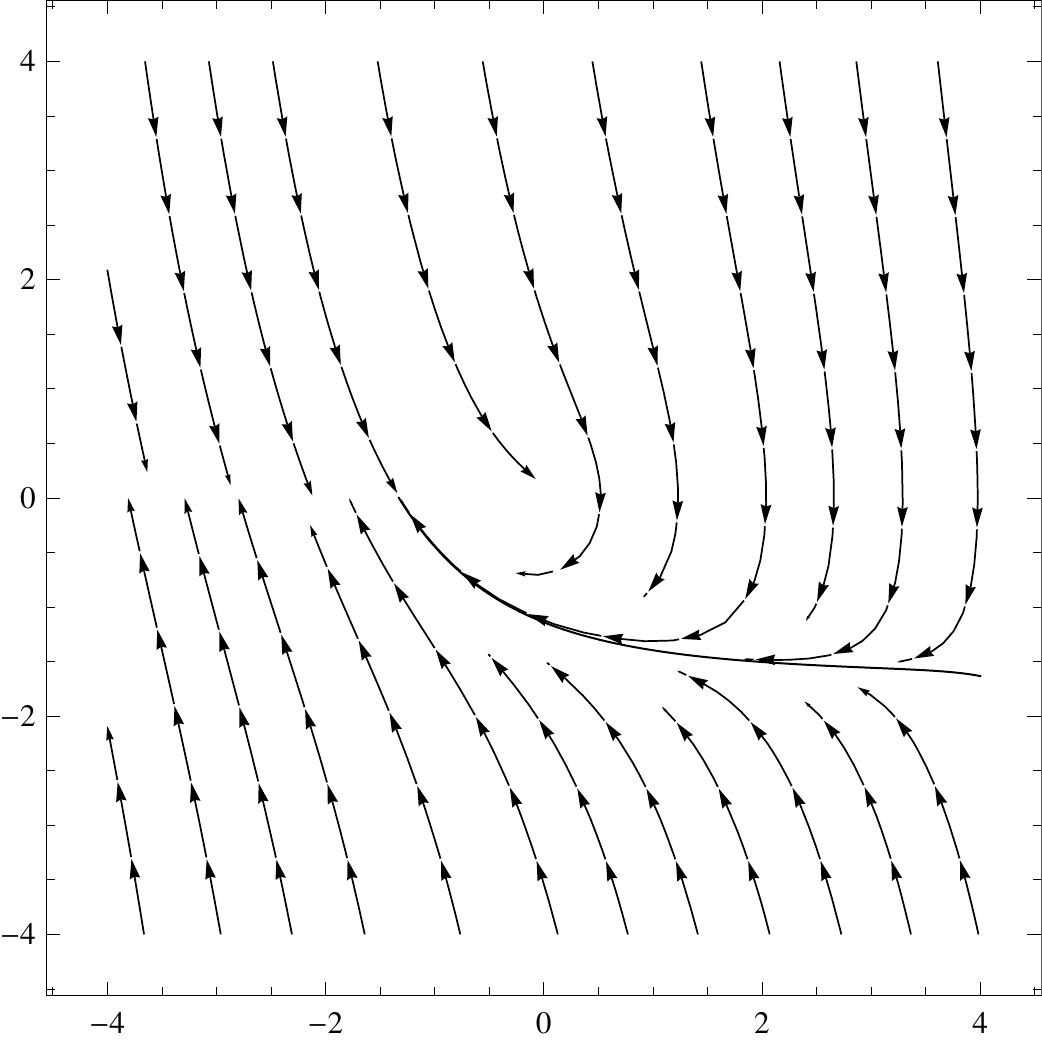} \\
& $\quad \frac{\displaystyle \varphi}{M_C}$ \\
\end{tabular}
}
\qquad
\subfloat[$\mu = 2 \mu_c$]{\label{phase4}
\begin{tabular}{cc}
\raisebox{3.5cm}{ $\frac{\displaystyle \dot\varphi}{m M_C}$ } & 
\includegraphics[width=7cm]{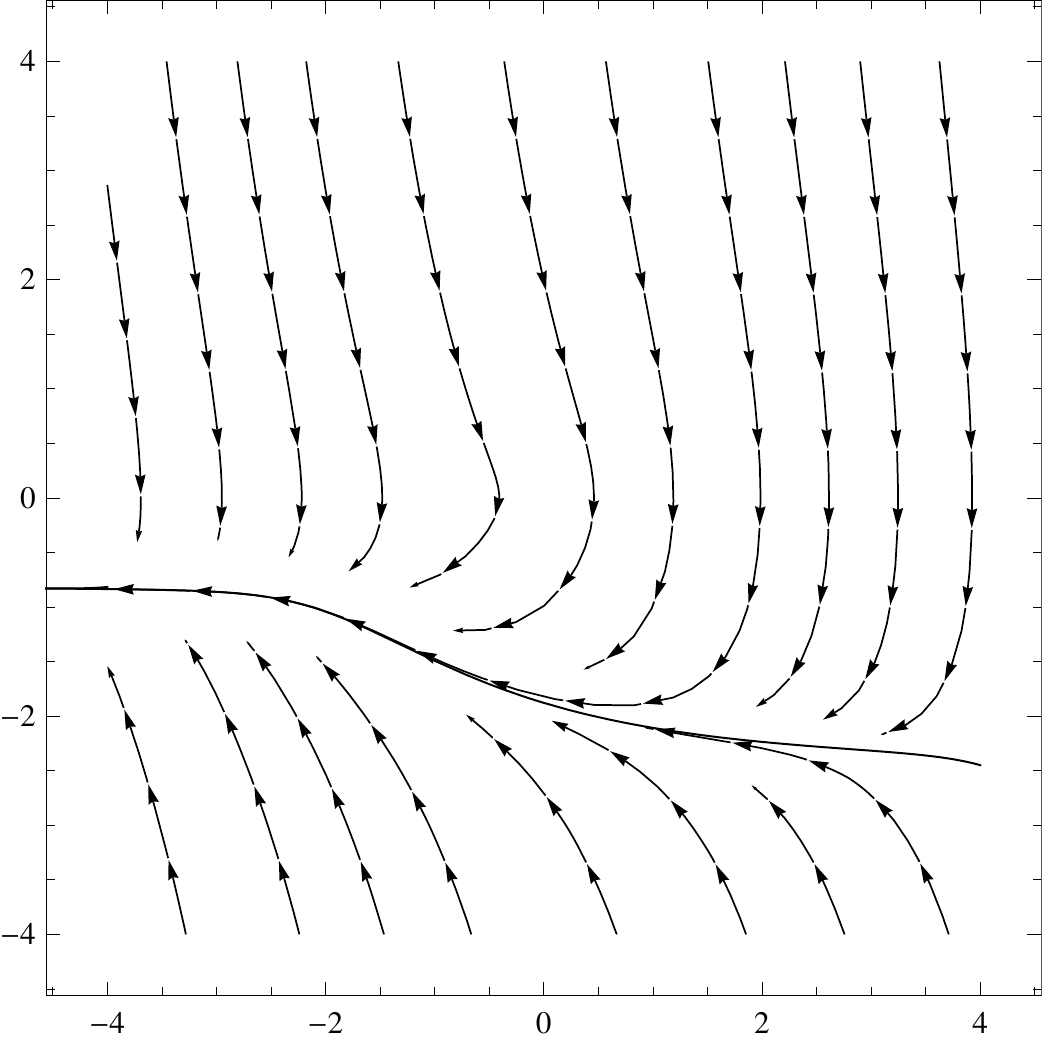} \\
& $\quad \frac{\displaystyle \varphi}{M_C}$ \\
\end{tabular}
}

\caption{ \label{phase-portrait}
Phase portraits of inflaton dynamics with the potential 
$V=\tfrac12 m^2\varphi^2 +\mu M \theta \varphi$, given by 
the dynamical system (\ref{dynamical_system_rescaled})
with $\mu>0$.  
\eqref{phase1} The result for standard slow-roll inflation. 
Solutions are attracted toward the slow roll solution, which appears as two almost-horizontal lines on the plot. At the end of the slow roll period, a reheating phase begins where the field undergoes damped oscillation about the minimum of the potential, which appears as an inward spiral in the phase portrait.
In \eqref{phase2} the basic picture remains the same, but the value 
of $\dot\varphi$ for the slow roll solutions is changed. 
The duration of the slow roll period is increased if $\varphi_i<0$
or decreased if $\varphi_i>0$.
In \eqref{phase3}, $\mu$ is at exactly $\mu_c$, so any 
configuration with $\varphi < 0$ and $\dot\varphi = 0$ is a fixed point,
corresponding to a de Sitter space-time.
When $\mu$ is increased beyond the critical value, as in \eqref{phase4}, the slow roll solution becomes unstable and $\varphi$ grows without bound.
Although only portraits with $\mu > 0$ have been shown, the system has combined $\mu \to -\mu$ and $\varphi \to -\varphi$ symmetry, allowing the phase portraits with $\mu < 0$ to be obtained from the phase portraits with $\mu > 0$ by a reflection through the origin.
}
\end{figure*}

The general behavior of this dynamical system for different values of $\mu/\mu_c$ 
is illustrated with phase portraits in Fig.~\ref{phase-portrait}. Plotted there is 
the flow on the $(\varphi, \dot \varphi)$ plane that is obtained when $\theta$ is
eliminated using the Friedmann equation. 
In terms of the dimensionless variables 
$\tilde\varphi = \varphi / M_C$ and $\tilde{t} = m t$, 
we can write the system in the form
\begin{equation} \label{dynamical_system_rescaled}
\frac{d}{d\tilde{t}}(\tvarphi, \tvarphi') = 
\left(\tvarphi',  - \tvarphi-\sqrt{\frac{3}{2} (\tvarphi'^2 + \tvarphi^2)}
 \left(\tvarphi' + \sqrt{\frac{2}{3}} \frac{\mu}{\mu_c} \right) \right),
\end{equation}
where the prime denotes derivative with respect to $\tilde t$.
The right hand side is the vector field plotted in the figure.

The number of e-folds of expansion
during inflation is
$N = \ln (a_f / a_i)=\int da/a$, 
where $a_i$ and $a_f$ are the 
scale factors at the beginning and end of inflation, respectively.	
$N$ can be computed in terms of the 
field values at the start and end of inflation as
\begin{equation}
N = \frac{1}{3} \int \theta\, dt 
= \frac{1}{3} \int \frac{\theta}{\dot\varphi}\, d\varphi.
\end{equation}
During slow-roll $\theta$ and $\dot\varphi$ are given by 
(\ref{slowroll-theta}) and (\ref{slowroll-varphi}), and
$\varphi$ has a fixed sign, so
\begin{equation}
N = \frac{\varphi_i^2 - \varphi_f^2}{4 M_C^2 \left(1 + \text{sgn}(\varphi_i) \mu / \mu_c \right)}.
\end{equation}
where $\varphi_i$ and $\varphi_f$ are the values of the inflaton field 
at the start and end of inflation, respectively.
As $\mu \to \mu_c$ one of the slow roll solutions approaches a 
de Sitter solution with constant negative $\varphi$, 
corresponding to the limit $N \to \infty$.
These solutions are illustrated in Fig.~\ref{phase3} on the negative
$\varphi$ axis.

The number of e-folds can be made arbitrarily
large compared to the $\mu = 0$ case, for fixed 
values of $\varphi_{i,f}$ and $M_C$, by taking $\mu \to \mu_c$.
However, the ratio $\mu/\mu_c$ is constrained
by the second inequality in (\ref{spin0stab}), which
is the condition that the spin-0 aether mode 
not grow exponentially in the long-wavelength limit.
Expressed in terms of $\mu_c$ (\ref{mu_c}), it yields
\begin{equation} \label{mu_over_mu_c1}
\frac{\mu^2}{\mu_c^2} \leq \frac{3 c_{13} + 3c_2} {2 + c_{13} + 3 {c_2}}.
\end{equation}
It follows 
that the limit $\mu \to \mu_c$ is achieved only by
$c_{13} \to 1$ or by $c_2 \to \infty$.
While the possibility that $c_{13}$ could be close to one
has not been ruled out, all the mode speeds 
\eqref{spin0speed}-\eqref{spin2speed} diverge
in this limit, suggesting that it is pathological.
Additionally, it is expected that a strong-field analysis of
binary systems will constrain $c_\pm$ to be
smaller than order unity.
As for the case where $c_2 \to \infty$,
as noted in Sec.~\ref{section:ppn}, 
$\varphi$ can be integrated out
so that the low-energy theory depends only on $c_2'$, given by (\ref{c2shift}).
It is possible to keep $c_2'$ small even while $c_2$ 
becomes large by taking $\mu / m$ large, though this 
requires a cancellation to occur between $c_2$ and $\mu^2/m^2$.

If the preferred-frame parameters 
$\alpha_1$ and $\alpha_2$ are constrained to
vanish exactly as in Sec.~\ref{section:ppn}, 
then as explained above, the constraint 
(\ref{mu_over_mu_c1}) is implied by the 
other constraints.
In this case, $c_2$ and therefore $\mu_c$ are 
determined by $\mu/m$ and $c_\pm$.
Then $\mu / \mu_c$ can be expressed as
\begin{equation} \label{mu_over_mu_c2}
\frac{\mu^2}{\mu_c^2} = \left(1 + \frac{2 m^2}{3 \mu^2} \frac{c_+ + c_- (1 - c_+)}{c_+ + c_-} \right)^{-1}.
\end{equation}
This approaches $1$ only if $\mu/m \to \infty$, or $c_+ \to 1$ and $c_- \to \infty$.
These are the same conditions as inferred above without having 
constrained $\alpha_{1,2}$ to vanish.

\subsection{Cosmological perturbations}

Finally, we briefly discuss the effect of 
inflaton-aether coupling on the spectrum of 
perturbations generated during inflation. 
In single-field slow roll inflation the 
primordial perturbations originate from 
quantum vacuum fluctuations of the coupled 
inflaton-metric mode. 
The introduction of the aether field leads 
to an additional spin-0 as well as a spin-1 mode, 
each with their own quantum fluctuations. 
If these modes are not directly coupled to matter and the background is exactly de Sitter, the aether modes decay exponentially \cite{Lim2004}, though they may be sourced during reheating by anisotropic stresses \cite{Li2007}. These perturbations may grow in power-law space-times for certain values of the parameters $c_{1,2,3,4}$ \cite{ArmendarizPicon2010}.

Introduction of scalar-aether coupling further modifies this scenario. There are two spin-0 coupled inflaton-aether-metric modes whose dispersion relation is nontrivial even in flat space. The generalization of this dispersion relation to curved space-time determines both the amplitude of vacuum fluctuations and the time at which the modes freeze. Moreover, the modified background equation of motion determines how the Hubble parameter changes with time, and therefore the time at which modes of different comoving wavelengths reach the Hubble radius.

A priori it is not clear whether the spin-0 modes will freeze or decay on superhorizon scales. This behavior is determined by how the modes couple to the background expansion, and is different for a scalar and the spin-0 part of a vector. 

\section{Summary} \label{section:conclusion}

We have considered Einstein-aether theory coupled to a scalar field via
a potential that depends on the local rate of expansion of space in the 
frame of the aether, $\theta=\nabla_a u^a$. This could be an effective field theory
description of Lorentz violating UV physics in the vacuum, for example, at
the scale of a fundamental cutoff. 
We have mostly focused on the lowest order term $\mu M\theta \varphi$, which
leads to breaking of time-reversal invariance, nontrivial dispersion, 
and a cosmological dynamics modified by a driving force on the
scalar proportional to $\theta$.

Although we are most interested in this model for the 
potentially observable effects of Lorentz violation on 
cosmology, we have strived to examine all of  
the theoretical and observational constraints that arise when
perturbing around a locally flat space-time.
We find that when the post-Newtonian parameters are matched to general relativity,
the combined constraints on the remaining free couplings $c_1$ and $c_3$ 
imposed by stability, positive energy density and absence of 
vacuum Cherenkov radiation, take the same form as in Einstein-aether theory,
except for the spin-0 Cherenkov constraint, which is relaxed (Fig.~\ref{parameter_range}). 
There is a single constraint on the new parameter $\mu$, 
the second inequality of (\ref{spin0stab}),
which is automatically satisfied when the PPN parameters 
match those of general relativity (\ref{mu_over_mu_max}).

An aether field uncoupled to matter does not affect 
the dynamics of homogeneous, isotropic, spatially flat 
cosmology except by a renormalization of Newton's constant 
$G \to G_C$ (\ref{G_c}). 
It turns out that 
the lowest order scalar-aether coupling $\theta\varphi$
contributes isotropic pressure and no energy density. 
Therefore it does not affect the Friedmann equation, 
but the driving force it adds to the scalar 
field equation can act either to oppose or accelerate 
the expansion of the universe in a slow-roll scenario, 
altering the number of e-folds of expansion. 
An unbounded increase
in the number of e-folds is possible, 
although this requires carefully chosen values of
the couplings in order to evade current constraints,
and may be ruled out by future constraints.
This illustrates the importance of considering all
constraints when building cosmological models.

A next step would be to derive the spectrum of primordial perturbations in
the model. This would involve generalizing the treatment of linearized perturbations from flat space to a general inflating background.
This differs from previous work since 
the two scalar modes---that of the inflaton and that of the aether---are mixed,
and because the slow-roll dynamics is modified. 
In particular, it would be interesting to determine how the 
spectral index and the tensor-to-scalar ratio 
depend on the coupling constants of the theory.

While we have focused on the coupling to a scalar inflaton, 
other types of coupling to the expansion are possible, 
allowing for a wide range of Lorentz-violating phenomena.
For example, any scalar operator could appear in the Lagrangian
multiplied by $\theta$, allowing effects to be switched off (or on)
as the expansion of the universe slows. In this context, 
it should be kept in mind that 
in the present universe the largest potentially observable 
values of $\theta$ are not due to the Hubble expansion,
but rather probably occur outside stellar mass black holes,
where $\theta\sim \mbox{km}^{-1}$.

\section*{Acknowledgments}

The authors thank Brendan Foster and Eugene Lim for helpful discussions.
This research was supported in part by the Foundational Questions Institute 
(FQXi grant RFP20816), NSF Grants No.~0601800 and No.~0903572, and by an NSERC PGS-D to WD.

\bibliographystyle{utphys}
\bibliography{aether-scalar}

\end{document}